\documentclass[aps, prx, 10pt, twocolumn, superscriptaddress,floatfix,longbibliography]{revtex4-2}

\usepackage{graphicx}
\usepackage{amsmath,amssymb,amsfonts,mathtools}
\usepackage{comment}
\usepackage{physics}
\usepackage[percent]{overpic}
\usepackage{braket}

\usepackage{bm}
\usepackage[usenames,dvipsnames, pdftex]{xcolor}
\usepackage[colorlinks, breaklinks, 
            linkcolor=OrangeRed,
            citecolor=RoyalBlue,
            urlcolor=RoyalBlue]{hyperref}

\usepackage{enumerate}
\usepackage[normalem]{ulem}

\newcommand{\NB}{{N_\mathcal{B}}} 
\newcommand{\rand}{r}
\newcommand{\nonrand}{nr}
\date{\today}

\begin{document}
	
\title{An SYK-inspired model with density-density interactions: spectral \& wave function statistics,  Green's function and  phase diagram
}

\author{Johannes Dieplinger}
\affiliation{Institute of Theoretical Physics, University of Regensburg, D-93040 Germany}
\author{Soumya Bera}
\affiliation{Department of Physics, Indian Institute of Technology Bombay, Mumbai 400076, India}
\author{Ferdinand Evers}
\affiliation{Institute of Theoretical Physics, University of Regensburg, D-93040 Germany}

\begin{abstract}
The Sachdev-Ye-Kitaev (SYK) model is a rare example of a strongly-interacting system that is analytically tractable. Tractability arises because the model is largely structureless by design and therefore artificial: while the interaction is restricted to two-body terms, interaction matrix elements are ``randomized" and therefore the corresponding interaction operator does not commute with the local density. Unlike conventional density-density-type interactions, the SYK-interaction is, in this sense, not integrable. We here investigate a variant of the (complex) SYK model, which restores this integrability. 
It features a randomized single-body term and a  density-density-type interaction. We present numerical investigations suggesting that the model exhibits two integrable phases separated by several intermediate phases including a  chaotic one. The chaotic phase 
 carries several characteristic  SYK-signatures including in the spectral statistics and the frequency scaling of the Green's function and therefore should be adiabatically connected to the non-Fermi liquid phase of the original SYK model. Thus, our model Hamiltonian provides a bridge from the SYK-model towards microscopic realism. 
\end{abstract}

\maketitle

\section{Introduction}
In recent years the Sachdev-Ye-Kitaev (SYK) models have received a fair amount of attention.\cite{maldacena2016remarks,kitaev2018soft,garcia2016spectral,polchinski2016spectrum} They represent a class of  Hamiltonians appropriate for modelling strongly interacting fermions:  
\begin{equation}
	\hat H_\text{cSYK}=\sum^{N_\mathcal{B}}_{i,j,k,l}J_{ijkl}c^\dagger_i c^\dagger _j c_k c_l - \mu \sum_{i} c^\dagger_i c_i, 
	\label{e1}
\end{equation}
where $c_i, c_i^\dagger$ denote the fermionic annihilation and creation operators at site $i$. The interaction kernel $J_{ijkl}$ represents a complex valued matrix with randomly distributed elements sampled from a Gaussian distribution with zero mean and variance $\langle J_{ijkl}^2\rangle=J^2/(2N_\mathcal{B})^3$ subject to the condition of  hermiticity. 
%

%
After choosing $J$ as the unit of energy, the Hamiltonian \eqref{e1} features only two parameters: the size of the single-particle basis $N_\mathcal{B}$
and the number of fermions $N$ or the chemical potential $\mu$, respectively. The temporal and spatial dynamics thus mediated is structureless in the thermodynamic limit in the sense that per time two particle-hole pairs are redistributed with the only constraint of particle-number conservation. 
Correspondingly, the spectrum follows predictions from random matrix theory (RMT) 
\cite{maldacena2016remarks, gu2020notes, kitaev2015simple} and the dynamics is quantum chaotic with details that depend on $N_\mathcal{B}$ and $\mu$.
\cite{maldacena2016remarks, behrends2020symmetry, garcia2018chaotic, haque2019eigenstate,kobrin2020many, garcia2017analytical,foot0}

{\bf Background.} The interest in the SYK-model derives from the fact that because of its conceptual simplicity it is exactly solvable at large $N_\mathcal{B}$, while despite its simplicity it reveals very interesting properties reviewed, e.g., in Refs.  \cite{maldacena2016remarks,gu2020notes, kitaev2015simple}.
One aspect of intrigue for many authors is the existence of a holographic gravitational dual to this model that captures, e.g., the long time fluctuations of the conserved quantities energy and particle number.
\cite{maldacena2016remarks,kitaev2015simple,gu2020notes, maldacena2016bound, maldacena1999large}. 
The relation to gravity embarks on the scale-free character of this model, which allows for very large  parametrical windows with power-law (``critical'') dynamics. Of particular interest to us is that the associated relaxation rate for low-energy excitations scales linear in temperature, $\sim T$; in quantum information terminology this implies that an SYK-system resembles a ``fast scrambler''. \cite{maldacena2016bound, maldacena2016remarks, kitaev2015simple}  

 In order to provide the model with an extra, competing scale, authors  \cite{lunkin2018sachdev,altland2019quantum,garcia2018chaotic,haque2019eigenstate,banerjee2017solvable} have complemented \eqref{e1} with a one-body term, i.e. SYK2. 
In Ref. \cite{altland2019syk} the new term was constructed so as to commute with the local occupation operator $\hat n_i {\coloneqq} c^\dagger_i c_i$. The resulting model then has an interpretation as a quantum dot with strong randomness; remnants of the SYK-physics have been identified, e.g., in the temperature dependent conductance. 
The authors of 
\cite{lunkin2018sachdev,altland2019quantum,garcia2018chaotic} 
implement a one-body term as random, infinite range hopping Hamiltonian. The competition between kinetic energy (Poisson-type level statistics) and the SYK-interaction (Wigner-Dyson type) 
leads to a transition from an integrable to a chaotic, non-Fermi liquid phase.

 Aspects of the SYK-model's phenomenology, i.e. fast scrambling, are reminiscent of a "strange metal phase", which does not support quasiparticle excitations and which is sometimes encountered in high-$T_c$ materials. \cite{proust2019remarkable,legros2019universal,varma2019linear,sachdev2007quantum,van2003quantum} 
 Therefore, the SYK-Hamiltonian has been proposed to model this phase. \cite{patel2019theory, altland2019sachdev, lunkin2018sachdev, parcollet1999non,cha2020linear} 
In order to compare in better detail,  transport properties such as the conductivity or resistivity are of great interest. Since they cannot be defined in the SYK-model proper due to its lack of spatial structure, extensions of Eq. \eqref{e1} have been proposed. 
One strategy has been to replicate the Hamiltonian $\eqref{e1}$ and then defining nearest-neighbors 
on a lattice of replicas \cite{patel2019theory, guo2019transport, chowdhury2018translationally}.  

It has been found that models of such or similar structure are able to reproduce Planckian resistivity and hence serve as a promising candidate for modelling strange metal electronic transport.

{\bf This work.} While the predictions for phenomenology may capture aspects of the strange metal phase, the specific form of models like \eqref{e1} is challenging to motivate: First, the Hamiltonian \eqref{e1} lacks the notion of space, which in more realistic models is imposed by the locality of physical interactions. 
Second, the interaction in \eqref{e1} is particle-number conserving, but not, in general, of the density-density type. The consequences of this latter observation we explore in this paper. 
Specifically, we study the  model Hamiltonian 
\begin{equation}
 \hat H_\text{tU} =\sum_{ij}^\NB t_{ij}c_i^\dagger c_j+\sum_{ij}^\NB U_{ij}\hat n_i \hat n_j,  
 \label{e2}
\end{equation}
where $\hat n_i{=} c_i^\dagger c_i$. 
In contrast to Eq. \eqref{e1}, the model \eqref{e2} has the attractive feature that its interaction term commutes with the local density $\hat n_i$ and thus shares a fundamental symmetry with physical interactions, in particular, the Coulomb interaction.
Other than this symmetry, we keep the model structureless as before with SYK: the constituting matrices $t_{ij}$ and $U_{ij}$ are taken to be a zero-mean complex Gaussian random number with variance $\langle t_{ij}^2\rangle{=}t^2/(64 \NB)$
and $\langle U_{ij}^2\rangle{=}U^2/(64 \NB)$; 
hermiticity requires  $t_{ij}{=}t_{ji}^*$ and $U_{ij}{=}U_{ji}{=}U_{ij}^*$. 
 As usual, the scale factor in the denominator ensures a volume-scaling of the total energy when taking the thermodynamic limit, see appendix  \ref{appendix:norm}. 

\begin{figure}[t]
    \centering
    \includegraphics[width=\linewidth]{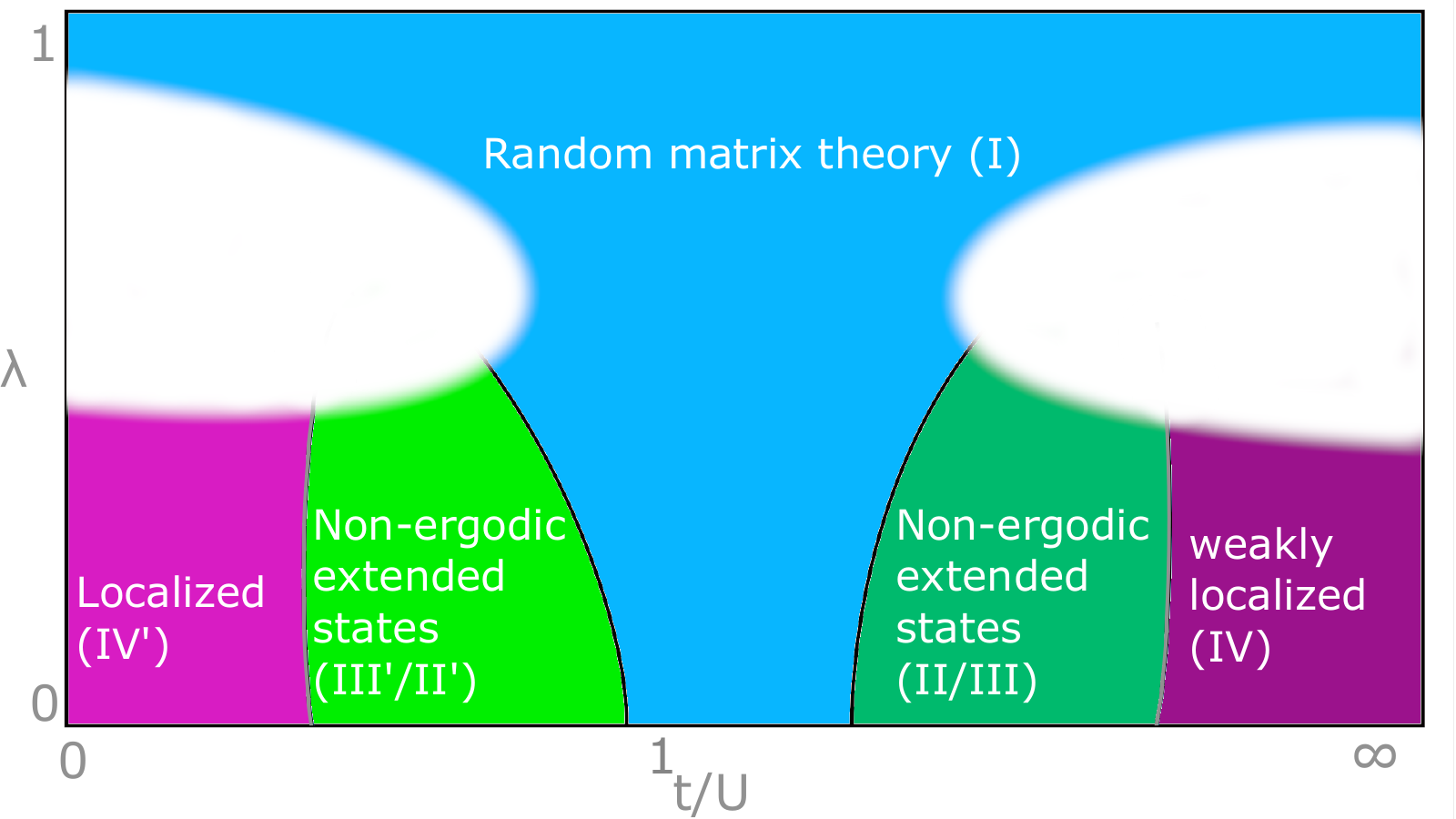}
    \caption{Phase diagram of model Eq. \eqref{e2}. The parameter $\lambda$, defined in Eq. \eqref{e7},  interpolates between the complex SYK model \eqref{e1},  $\lambda{=}1$, and  \eqref{e2}, $\lambda{=}0$.  The regions I, II/III and IV have been identified according to the classification given in Ref. \cite{monteiro2020minimal}
     Black phase boundaries denote phase transitions, while grey boundaries are expected to collaps with the left and right edges in the thermodynamic limit. Blank regions are unknown. }
    \label{f0}
\end{figure}

The properties of the model \eqref{e2} will be explored in the following. We give a brief account of our findings. 
In Eq.  \eqref{e2} each term by itself is integrable in a many-body sense: While the bare interaction term conserves the local occupations, the kinetic term proper conserves the occupations of single-particle eigenstates. Due to this trivial integrability one finds two fixed points, each with Poisson-type level spacing at $t{=}0$ and $U{=}0$, respectively. 
These fixed-points, - which presumably become unstable already when $t$ and $U$ differ from zero, - are
separated by a region exhibiting a Wigner-Dyson-type level statistics, 
see Fig. \ref{f0}, with several subphases. 
The properties of these phases we characterize in terms of the inverse participation ratio in Fock-space and the spectral form factor. While  the three sectors I, II/III, IV are naturally identified with phases found earlier by \textcite{micklitz2019nonergodic,monteiro2020minimal}, the sector IV' differs from IV and has not been described before. 

Finally, we calculate the imaginary frequency Green's function and compare it with the well-known conformal SYK results in the infrared and large $N_\mathcal{B}$-limit:  the frequency scaling of  \eqref{e2} is consistent with an SYK-type behavior. While the available system sizes are too small to make definite predictions for the thermodynamic limit, our results still suggest that both phases are adiabatically connected and therefore should be identified with each other. In this sense and similar to Ref. \cite{altland2019quantum} our work tightens the link between the SYK-model and more realistic microscopic ones. 

\section{Method \label{s2}} 

In order to calculate statistical and dynamical properties we diagonalize Hamiltonian \eqref{e2} exactly for a finite number of fermionic sites $N_\mathcal{B}=12,16,20$.
\cite{foot1}
The Hamiltonian is written in Fock space in the occupation number basis of the operators $\hat{n}_i$. Recalling particle number conservation, the Hamiltonian matrix can be brought to block diagonal form, such that the effective Fock space dimension reduces considerably for the individual blocks associated with a specific filling. In this work we always choose a quarter filling.
\cite{foot2}

\begin{figure*}[t!]
    \centering
	\includegraphics[width=\linewidth]{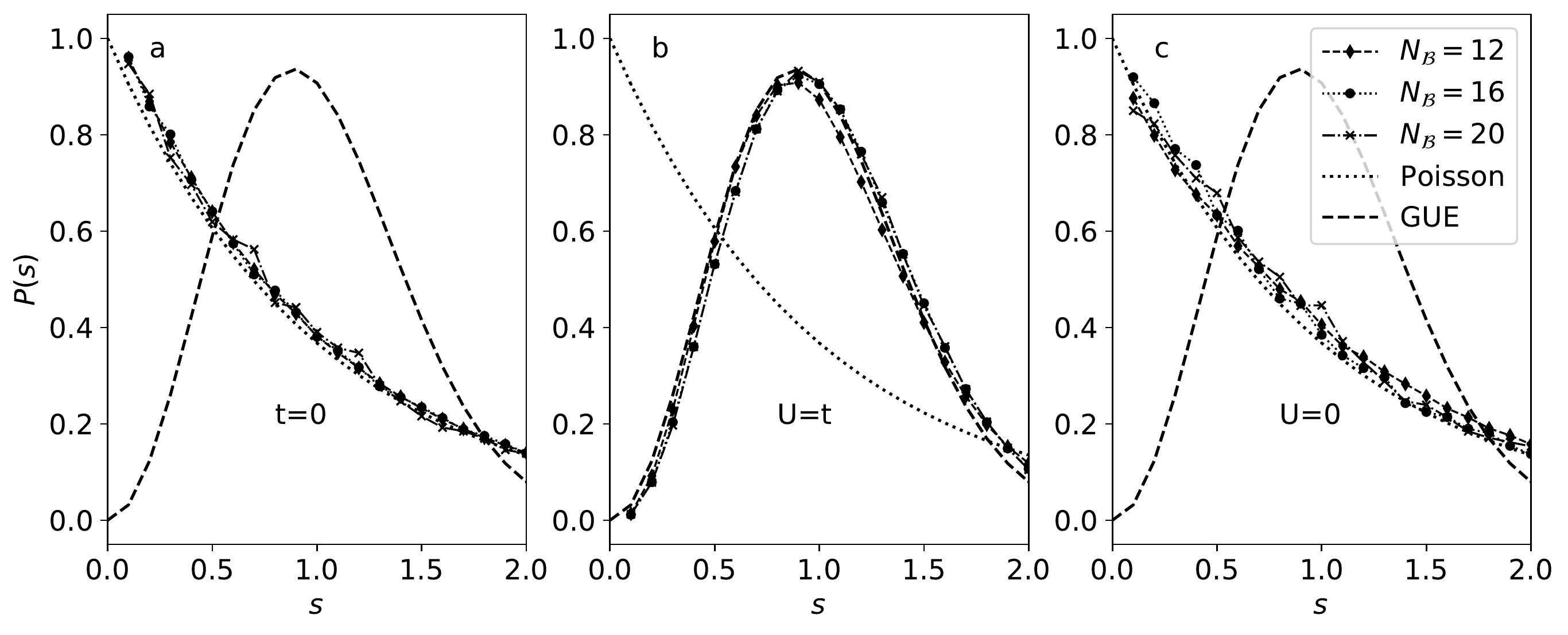}

	\caption{Level spacing statistics in phases IV' (a), I (b), and IV (c); nomenclature defined in Fig. \ref{f0}. Dashed curve indicates GUE prediction, dotted line the Poisson limit. (number of disorder realizations: 5000 for $N_\mathcal{B}=12$, 200 for $N_\mathcal{B}=16$ and 20 for $N_\mathcal{B}=20$. The filling is $1/4$.) The plot shows that despite the limited system sizes an assignment of localized vs delocalized to the respective phases is justified.}
	\label{f1}
\end{figure*}

\section{Computational Results}
The close and prominent relation between the SYK model and random matrix theory (RMT) plays a crucial role for the holographic duality to black holes in two-dimensional 
 gravity.
\cite{maldacena2016remarks}
We therefore begin the outline of our computational results with the level spacing statistics $P(s)$, 
which is a hallmark of RMT-related physics. 
A subsequent analysis of level spacing ratios serves for a first mapping of the different phases involved and their respective boundaries. 
As it turns out, $P(s)$ and the level spacing ratios are blind to a subspace structure, which however readily emerges after consulting the spectral form factor $K(T)$. In fact, the form factor also can discriminate the localized phases IV and IV' from each other. 
The substructure of the delocalized phases also reveals in the inverse participation ratio, which we investigate subsequently. 
We conclude with a study of the frequency dependency of the Green's function.

\subsection{Level spacing statistics}
\subparagraph*{Distribution function $P(s)$.}

The level-spacing distribution function is defined as 
\begin{equation}
    P(s) = \overline{\delta(s-s_l),} 
    \end{equation} 
where $s_l=(E_{l}-E_{l-1})/\Delta$ denotes the spacing between two neighboring eigenvalues of the Hamiltonian \eqref{e2} in units of the corresponding mean level spacing $\Delta$. We take pairs of energies close to the center of the spectrum including 60\% of the spectrum. In Fig. \ref{f1} b we display the result for the model \eqref{e2} at quarter filling and $t=U$, i. e. in phase I. The form of $P(s)$ is consistent with RMT-expectations, i.e.  the Wigner-Dyson distribution of the Gaussian unitary ensemble,  \cite{garcia2018chaotic,behrends2020symmetry}
\begin{equation}
    P_\text{GUE}(s)=\frac{32}{\pi^2}s^2e^{-\frac{4}{\pi}s^2}. 
    \label{e5} 
\end{equation}
We thus observe a clear signature of quantum chaotic dynamics; the system under investigation is not integrable at $t=U$. For integrable phases, such as IV and IV', an absence of level repulsion is expected reflecting in the Poisson  statistics $P_\text{P}(s)=e^{-s}$, see Fig. \ref{f1} a, c. 

The RMT-behavior observed in phase I is consistent with findings in the SYK model \eqref{e1} -- away from half filling. 
\cite{foot3}
 Our  model \eqref{e2}  complies with this trend towards the  Wigner-Dyson limit, Eq. \eqref{e5} suggesting that phase I and the SYK-phase are adiabatically connected.

\subparagraph{Level spacing ratios $r(s)$.}
In order to map out the phase boundaries, we employ the level spacing ratios \cite{cuevas2012level,garcia2018chaotic}, defined as the
\begin{equation}
	r_l=\frac{\min(s_l,s_{l+1})}{\max(s_l,s_{l+1})}. 
	\label{levels}
\end{equation}
 The double average of $r_l$ first over all pairs $(s_l,s_{l+1})$ per sample and second over the ensemble of samples defines the parameter $\overline{r}$, which is capable of distinguishing, e.g., different RMT-ensembles, as well as Poissonian level statistics. The latter corresponds to a value of $\overline{r}\approx0.38$, while GUE statistics implies  $\overline{r}\approx0.599$ \cite{atas2013distribution,cuevas2012level,garcia2018chaotic}.

\begin{figure}[t]
\includegraphics[width=\linewidth]{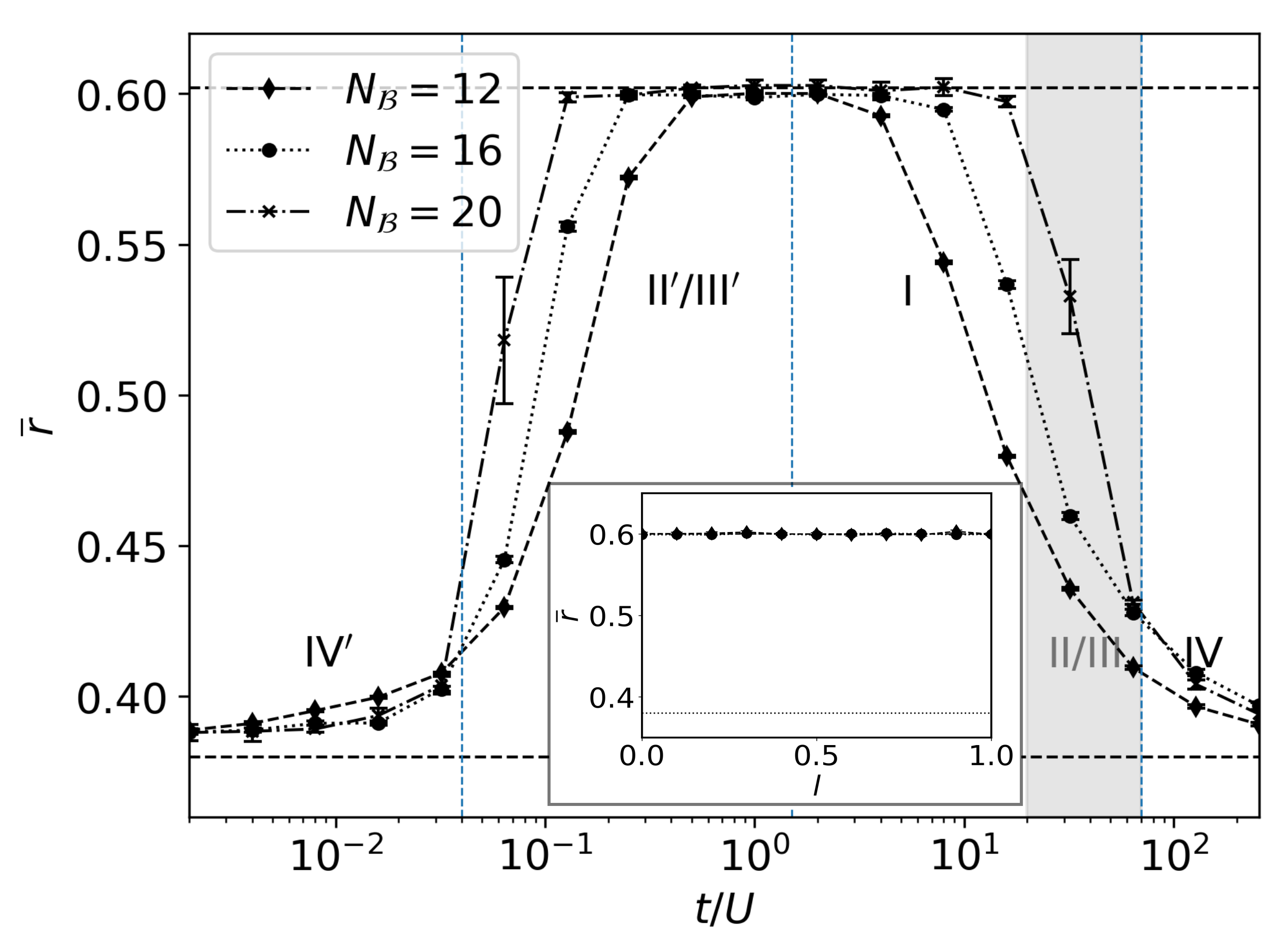}
    \caption{Phase transition between localized (Poisson-type) and delocalized (Wigner-Dyson-type) level statistics. The dashed lines indicate Poissonian versus GUE behavior. \cite{atas2013distribution} The nomenclature I, II/III and IV follows Ref. \cite{micklitz2019nonergodic}; the labeling II'/III' and IV' is a generalization emphasizing the analogy. The different nature of the subphases I and II/III is not resolved with $\overline{r}(\lambda)$.
	Inset: Adiabatic connection of the level spacing statistics between Eq. \eqref{e1} and \eqref{e2}, setting $U=t=J=1$ with 5000/200/5 disorder realizations for $N_\mathcal{B}=12,16, 20$. 
	}
	\label{f2}
\end{figure}

We determine the ratio $\overline{r}$ in the parameter regime $10^{-3}\lesssim t/U \lesssim 200$, see Fig. \ref{f2}. 

Two transitions are seen,  which separate 
the Poissonian boundary phases at $t{=}0$ (IV') and $U{=}0$ (IV) from a non-integrable region in between. 

The transition near $U{=}0$ into IV is similar to the quadratic complex SYK model studied in Refs. \cite{liao2020many,winer2020exponential}. %
Integrability of the randomized all-to-all hopping at $U{=}0$ has an expression in terms of a large $SU(2)$ symmetry group, which tends to destroy level correlations. \cite{winer2020exponential,liao2020many} 

In order to gather further evidence for identifying the chaotic phase seen in Fig. \ref{f2} with the SYK-phase embodied in Eq. \eqref{e1}, we extend the analysis of the level spacing ratios. We define a family of Hamiltonians that interpolates between the two models, Eq. \eqref{e1} and \eqref{e2}, with a parameter $\lambda\in[0,1]$: 
\begin{equation}
    \hat{H}(\lambda)=\lambda\cdot \hat{H}_\text{cSYK} + (1-\lambda)\cdot \hat{H}_\text{tU}.
    \label{e7} 
\end{equation}
The level spacing ratios will, in general, depend on the family parameter, $\overline r(\lambda)$, taking a critical value, which differs from the RMT-value, if a phase-transition is crossed. 

The inset of Fig. \ref{f2} displays the ratio $\overline{r}(\lambda)$ all along the line in parameter space connecting  Hamiltonian \eqref{e1} at $\lambda{=}1$ to Hamiltonian \eqref{e2} at $\lambda{=}0$. As is seen, the ratio is insensitive to $\lambda$, 
i.e., there is no indication of a phase transition - at least away from half filling and within the "range of visibility" of $\overline{r}(\lambda)$. The result strengthens the case  that the chaotic phase of model \eqref{e2} can indeed be identified with the non-Fermi-liquid phase of the SYK model. 

\begin{figure}[t]
	\includegraphics[width=\linewidth]{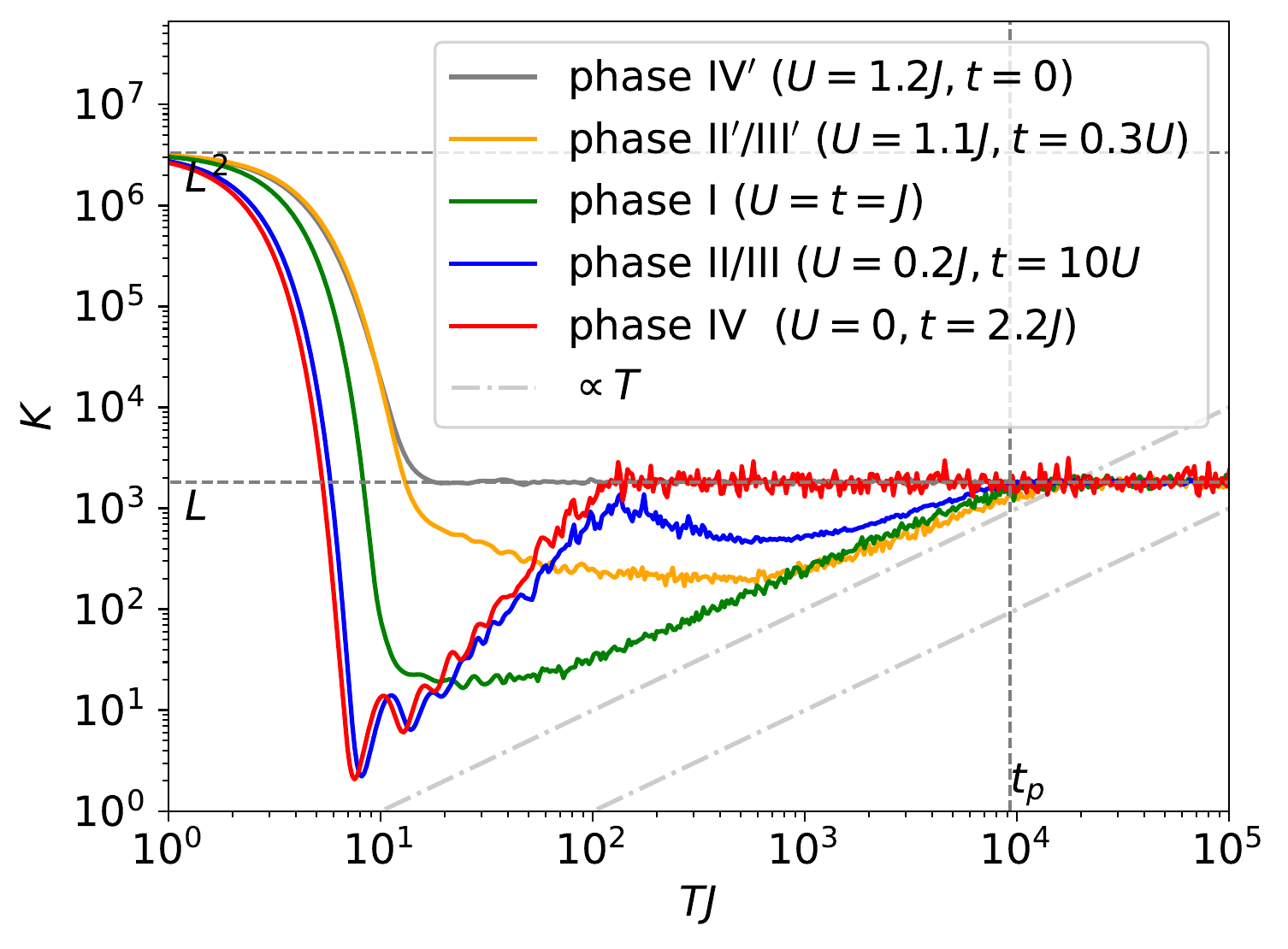}
	\caption{Spectral form factor with its shape characteristic for the different phases IV', II'/III', I, II/III and IV. The dashed line indicates the behavior $K(T)\propto T$ as is appropriate for the RMT-type behavior also observed within SYK. Within region IV ($U{=}0$) we see a growth stronger than $T$. With our system sizes, the exponential behavior expected for integrable systems cannot be resolved.\cite{winer2020exponential,liao2020many} (Parameters: $N_\mathcal{B}=16$, 1000 disorder realizations, $t_p=5L/J$)}
	\label{fig:SFF}
\end{figure}

\subparagraph{Spectral form factor.}
Also the spectral form factor,
$K(T)$, 
discriminates different phases from each other - and, as it turns out - with a sensitivity superior to $P(s)$ and $\overline{r}(\lambda)$; formally, 
\newcommand{\ci}{\mathfrak{i}}
\begin{equation}
    K(T)=\overline{ \mid Z(\beta+\ci T)\mid^2},
\end{equation}
where  $Z(\beta {+}\ci T)$ is the partition function taken at (inverse) temperature $\beta$ and observation time $T$. The form factor $K(T)$ can be interpreted as a Fourier transformed correlator between energy levels. 
It is an established alternative descriptor of the level statistics and, hence,  quantum chaos. 
In our simulations we focus on 
 infinite temperature to probe homogeneously the entire spectrum.
\cite{brezin1997spectral,cotler2017black}  Then, a linear "ramp"-type feature characteristic of quantum chaotic systems is expected to appear in $K(T)$; it reveals itself in Fig. \ref{fig:SFF} 
at times $TJ\gtrsim \mathcal{O}(100)$, e.g., in phase I 
and it also emerges in the SYK model proper.\cite{cotler2017black,altland2018quantum,saad2018semiclassical} 
We calculate the spectral form factor for different relative interaction strengths $t,U$ such that the bandwidth fits the the SYK band width set by $J$ in Eq. \eqref{e1}. Therefore, the energy scales of the two models as well as the different parameter settings of Eq. \eqref{e2} can be directly compared. 
In the case of the quadratic model, phase IV ($U{=}0$), an exponential type of ramp has been reported for $K(T)$, based on analytical arguments and a numerical study. \cite{liao2020many,winer2020exponential}
In Figs. \ref{fig:SFF} 
we confirm a ramp with very strong growth, e.g. in $TJ\gtrsim \mathcal{O}(100)$ for phases II/III and IV, increasing significantly faster than linear in $T$. In our model, Eq. \eqref{e2}, the exponential growth cannot be identified, however, in the hopping-dominated regime at $U{=}0$, i.e. in region IV - at least not within the system sizes at our disposal.

Interestingly, the purely interacting model, i.e. region IV' ($t{=}0$), does not exhibit a ramp-like behavior, even though it is integrable as is region IV. Thus in contrast to the level spacing ratio $P(s)$, the form factor $K(T)$ distinguishes the two integrable phases, IV and IV', from each other.

This finding is consistent with  \textcite{prakash2020universal}, where above the localization transition triggered by a disordered Fock space diagonal term with completely uncorrelated energies, the linear ramp of an originally chaotic Hamiltonian vanishes. 

In region I ($t{\approx}U$) we see in Fig. \ref{fig:SFF} an approximately linear ramp consistent with SYK and expectations from RMT and quantum chaos. In agreement with the SYK and RMT results the plateau assumes of $L=\binom{N_\mathcal{B}}{N_\mathcal{B}/4}$ -- the number of energy levels -- at the time $t_p\sim L/J$. 

We note as a net outcome of this section that the model \eqref{e2} has a structure in phase space substantially richer as compared to the SYK-model proper, 
\eqref{e1}. While this structure does not reveal in the simplest descriptors of the level statistics, i.e. $P(s)$ and $\overline{r}(\lambda)$, it becomes manifest in the spectral form factor $K(T)$. 

We add two remarks on further research that our observations could motivate.
(a) The regions II/III and II'/III' exhibit their own individual and complex characteristics in the spectral form factor $K(T)$. Our preliminary data indicates a strong evolution of the shape of $K(T)$, seen in Fig. \ref{fig:SFF} in an intermediate time regime between $10\lesssim TJ \lesssim 1000$. The detailed nature of this regime requires further studies.

(b) An interesting extension of the analysis of $K(T)$ concerns the low-temperature regime, $J\beta\gg 1$. Indeed, studies of other SYK-derived Hamiltonians that also have been equipped with a 1-body hopping term have indicated the possibility for a cross-over from a chaotic, incoherent metal regime at $\beta=0$ to a (renormalized) Fermi liquid that occurs at low-enough temperatures. \cite{Song,parcollet1999non} 
The finite temperature behavior of our model \eqref{e2} has not been investigated yet and warrants a separate investigation. It will interesting to see if also in our model \eqref{e2} the high-temperature non-FL phase crosses over to a quasi-particle dominated regime at lowest temperatures.

\subsection{Wave function statistics}
The substructures I and II/III of the chaotic region have been analyzed before by
\textcite{micklitz2019nonergodic,monteiro2020minimal}. For identification these authors  employed the inverse participation ratio (IPR), which - for a given Fock-space basis - discriminates between localized states and also extended states with a different degree of delocalization.  \cite{monteiro2020minimal}
The formal definition reads  
\begin{equation}
I_q=\sum_{\nu} \overline{ \mid\braket{\nu\mid \Psi}\mid ^{2q}}
\end{equation} 
where the sum is over the Fock-space basis $\ket{\nu}$, see Sect. \ref{s2}.
\cite{foot4}
Here, the overline indicates a double average over eigenvectors $\ket{\Psi}$ from the bulk of the spectrum of a given many-body Hamiltonian and over the disorder ensemble. 
Trivially, for completely localized wave functions we have $I_q=1$ while for delocalized ones we will obtain the Porter-Thomas (PT) distribution 
\begin{equation}
	I_q=q!\mathcal{V}^{1-q},
	\label{e10}
\end{equation}
where $\mathcal{V}=\binom{N_\mathcal{B}}{\NB/4}$ denotes the Fock space volume. 
Eq. \eqref{e10} indicates that the statistics of wave-function components resembles independently distributed Gaussian variables. 

\begin{figure}[t]
	\includegraphics[width=\linewidth]{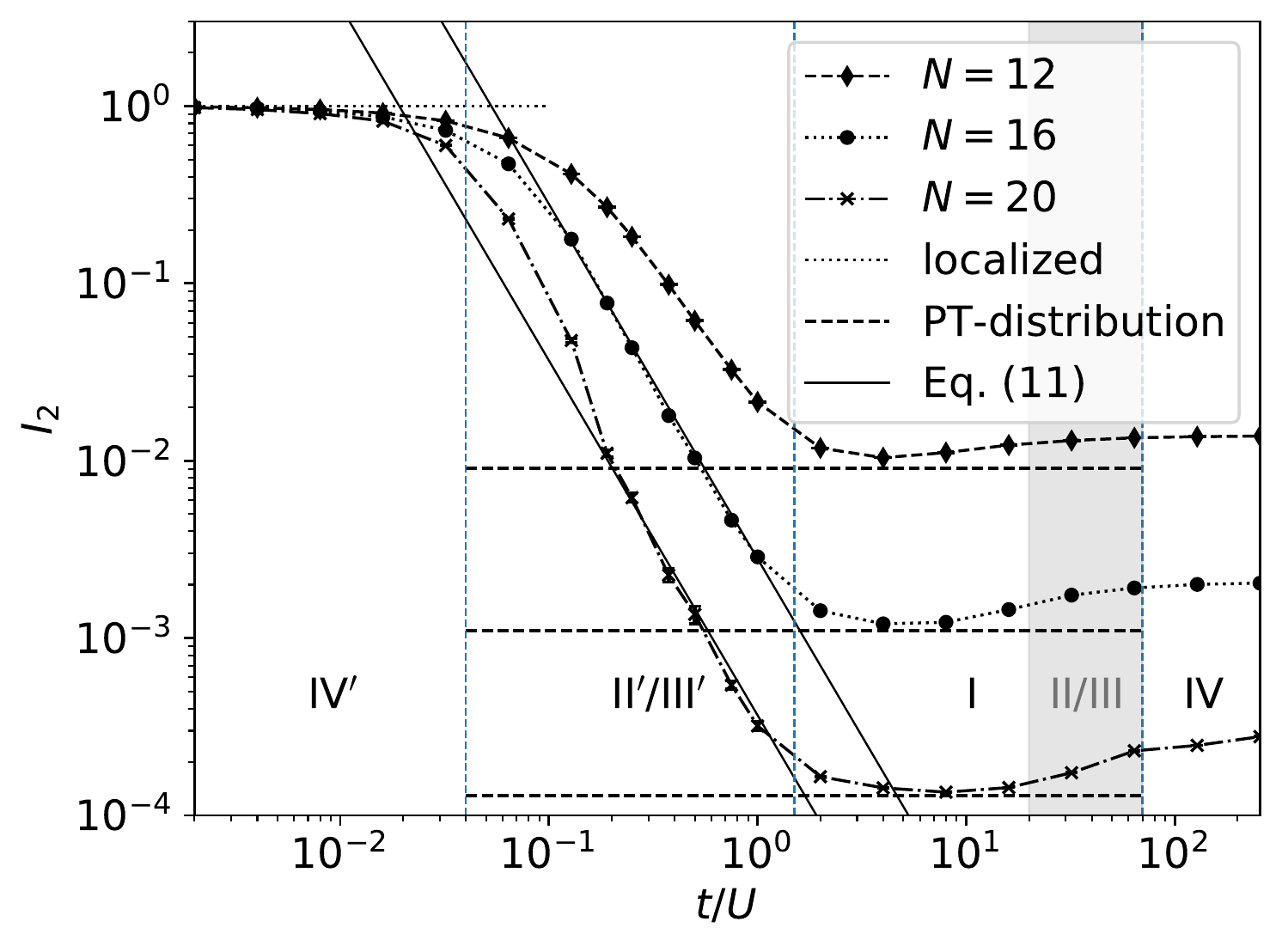}
	\caption{Inverse participation ration $I_2(t/U)$ for the phase-diagram Fig. \ref{f1} at $\lambda=0$ and system sizes $N_\mathcal{B}=12,16,20$.  The dashed lines indicate genuine RMT behavior for a given systems size.\footnote{Baselines have been obtained by diagonalizing random unitary hermitian matrices of appropriate size.} Solid line indicates the NEE predictions, Eq. \eqref{e10a}, with the constant $c=0.5$. 
	(Parameters: disorder average is performed over $5000/200/5$ samples for $N_\mathcal{B}=12,16,20$  realizations.) 
	}
	\label{f5}
\end{figure}

In Fig. \ref{f5} we
depict our computational result for $I_2$ and model \eqref{e2}.
When averaging only wave functions from the center of the spectrum, i.e. the central fifth of the energy levels, have been included.
We observe several qualitatively different regimes. 

{\bf Region IV'.}  In the absence of hopping ($t{=}0$) wavefunctions are localized in our chosen  Fock-basis reflecting the fact that the interaction is of the density-density type and therefore diagonal in the basis chosen; $I_2\approx 1$, correspondingly. The finding is consistent with integrability and our previous results derived from the level statistics. 

{\bf Region I.}
In this regime, $t\gtrsim U$, RMT holds and the IPR approaches the Porter-Thomas result, Eq. \eqref{e10}, with increasing system size $N_\mathcal{B}$.  

{\bf Region IV.} The model is integrable also in the non-interacting limit $U{=}0$. However, unlike it is the case with IV', the localized character of wavefunctions in Fock space does not reveal with the IPR in Fig. \ref{f5} due to the choice of our basis. For resolving this behavior the proper choice of basis is given by the eigenstates of $t_{ij}$ in Eq. \eqref{e2}. 

{\bf Region II/III.}
Refs. \cite{micklitz2019nonergodic,monteiro2020minimal} suggest that there is an  intermediate phase (with two subphases II and III) that exhibits extended states, which do not fill all of the available Fock space (``non-ergodic''). In this non-ergodic extended (NEE) phase the level spacing statistics $P(s)$ is of the random matrix type. However in contrast to phase I, $I_2$ does not accept the Porter-Thomas form independent of $t/U$ in II/III, in the limit of large $N_\mathcal{B}$. As was already the case with IV, our choice of basis does not allow us to reveal this feature in Fig. \ref{f5}. 

{\bf Region II'/III'.}
However, in our model there is a dual phase II'/III', which separates the phase I from a second integrable point IV'. While the form-factor shows that this point is not to be identified with IV, it nevertheless allows for intermediate phases II'/III' analogous to II/III with IV. 
Instead, the IPR $I_2(t/U)$ is seen in Fig. \ref{f5} to exhibit 
a strong variation with its argument. It is a manifestation of the fact that wave functions occupy a small but non-zero fraction of the Hilbert space; they are neither localized nor do they fully spread within the entire Fock space. 

In Refs. \cite{micklitz2019nonergodic,monteiro2020minimal} the IPR $I_2(t/U)$ 
is predicted to follow a power law, $I_2(t/U) \propto (t/U)^a$; the exponent $a$ is integer, either $a{=}1$ (in III) or $a{=}2$ (in II).\cite{monteiro2020minimal} 
Our data confirms to this expectation with respect to III and $a{=}2$, see Fig. \ref{f5}. The phase II with $a{=}1$ we do not resolve here, possibly due to restrictions in the available system size. 
For further analysis we adopt a formula derived analytically in  \cite{monteiro2020minimal}. It predicts the inverse participation ratio to follow a power-law 
\begin{equation}
 I_2=c\cdot8\sqrt{N_\mathcal{B}}\frac{U^2}{\pi t^2 \binom{N_\mathcal{B}}{N_\mathcal{B}/4}}. 
 \label{e10a}
\end{equation}
While the parametric scaling is expected to carry over also to our model, the prefactor $c$ is left unspecified; Fig. \ref{e5} suggests $c\approx0.5$.
%

\subsection{Green's function}
The RMT-phase I of model \eqref{e2} is of central interest, because it may connect to the strong-coupling phase of the SYK model \eqref{e1}. We here continue our investigation of the phase I to provide further evidence supporting this adiabatic connectivity.  
Indeed, we already have demonstrated that the model \eqref{e2} reproduces the characteristic statistical properties of the SYK model \eqref{e1} in the chaotic phase and also the detailed structure of subphases that appear in SYK-derivatives. \cite{micklitz2019nonergodic, monteiro2020minimal}
We now show that the models \eqref{e1} and \eqref{e2} may indeed share the characteristic power law structure of the Green's function, $G(\ci\omega)\propto \omega^{1/2}$ reflecting the conformal symmetry \cite{maldacena2016remarks,kitaev2015simple}, which is expected to hold in the frequency window 
\begin{equation}
    \frac{J}{N_\mathcal{B}}\ll\omega \ll J.
    \label{e11} 
\end{equation}
\begin{figure}[t]
    \includegraphics[width=\linewidth]{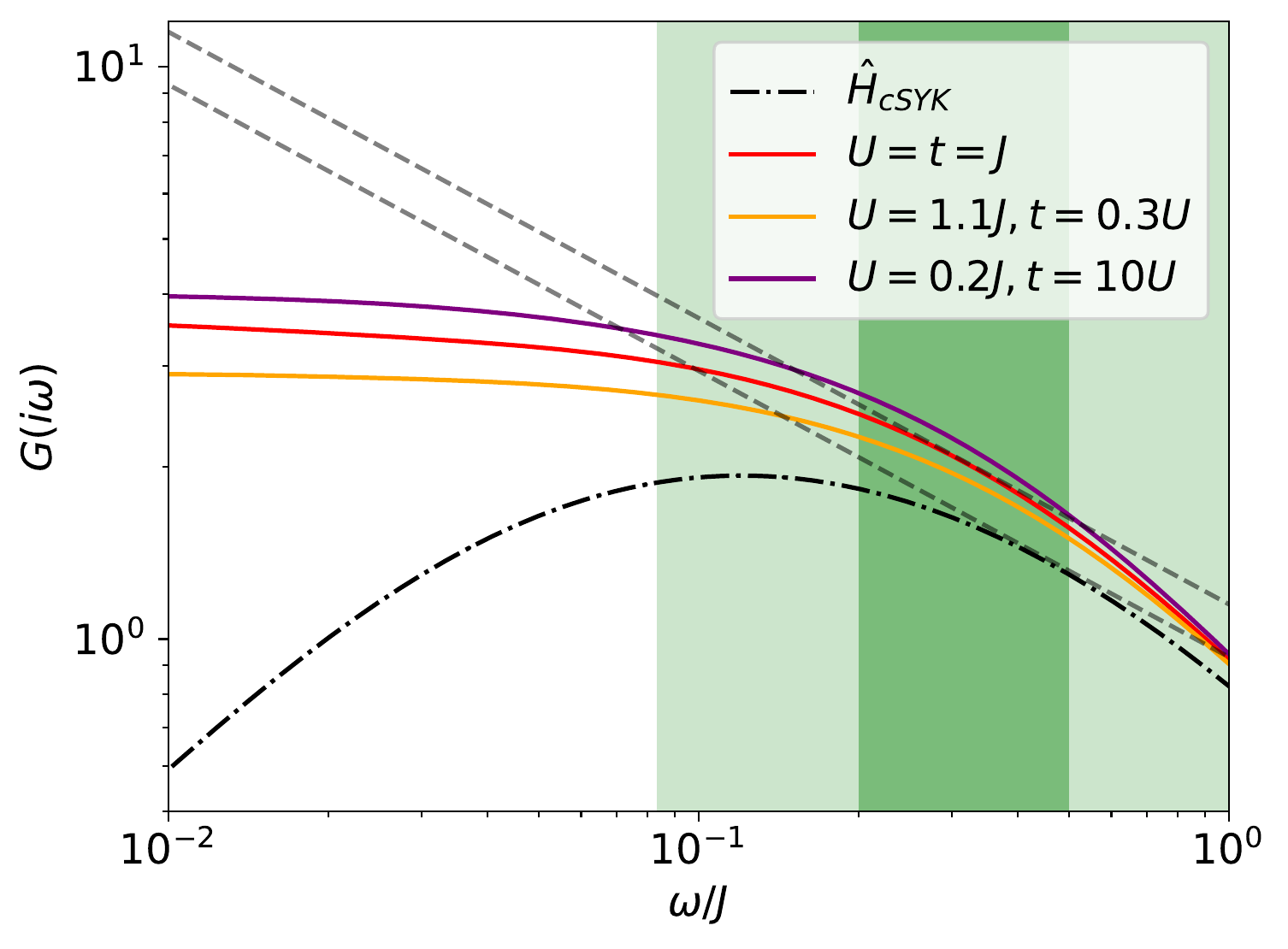}
    \caption{ Matsubara-Green's function of the model \eqref{e2} for  $N_\mathcal{B}=12$. The lightly shaded area, $J/N_\mathcal{B}{<}\omega{<} J$, indicates the conformal window, \eqref{e11}; the dashed line represents the corresponding power law scaling $G(\omega) {\propto} \omega ^{-1/2}$ scaling. Parameter settings of Hamiltonian \eqref{e2} have been rescaled to ensure comparability between the respective models. 
    }
    \label{f6}
\end{figure}

Formally, the imaginary frequency Green's function is defined as the Fourier transform of
\begin{equation}
    G(\tau)=G_i(\tau)=\langle \comm{c_i^\dagger(\tau)}{c_i(0)}\rangle,
\end{equation}
where the expectation value is taken over the ground state, we have a Lehmann representation
\begin{equation}
    G(\ci\omega)=\sum_n \frac{\bra{0}c_i\ket{n}\bra{n}c_i^\dagger \ket{0}}{\ci\omega +E_0-E_n}+\frac{\bra{0}c_i^\dagger\ket{n}\bra{n}c_i \ket{0}}{\ci\omega -E_0+E_n}.
    \label{e13} 
\end{equation}
Here, $\ket{n}$ labels the exact eigenstates of the Hamiltonian and $E_n$ the corresponding eigenenergies. $\ket{0}$ is the groundstate. We evaluate \eqref{e13} by exact diagonalization of the Hamiltonians \eqref{e1} and \eqref{e2}. 
The result is depicted in Fig. \ref{f6}. As one might have expected, within the relevant frequency window the shape of $G(\ci \omega)$ as obtained for \eqref{e2} is indeed roughly consistent with the SYK result \eqref{e1}. In both cases, the critical window of frequencies is two narrow, however, in order to clearly identify a power law. That finite size effects could indeed be significant is seen from the fact that the scaling observed in the neighboring phases II/III and II'/III' is similar to the one seen in I. Since these phases are not of the RMT-type proper, a scaling with $\omega^{1/2}$ is not necessarily expected, here.


\begin{table}[b]
\begin{tabular}{p{1.65cm}|p{1.45cm}|p{2.85cm}|p{0.5cm}|p{0.65cm}|p{0.5cm}}
    \hline \hline 

 References & model & $\alpha, \beta$ & $\tilde{t}_{ij}$ & $\tilde{U}_{ij}$ & $\epsilon_i$\\
\hline 

Ref.~\cite{kitaev2015simple,maldacena2016remarks} & majorana & $\alpha,\beta{=}0$ & 0 & SYK & \nonrand \\

Ref.~\cite{gu2020notes,fu2016numerical} & fermion & $\alpha,\beta{=}0$ & 0 & SYK & \nonrand \\

Ref.~\cite{monteiro2020minimal,garcia2018chaotic,haque2019eigenstate,lunkin2018sachdev} & majorana & $\alpha,\beta{=}0$ & \rand & SYK & \nonrand \\

\hline

This work & fermion & $\alpha, \beta {=} 0$ & \rand & \rand & \nonrand \\

\hline

Ref.~\cite{BurinPRB15} & spin & $\alpha,\beta \in (d,3)$ & \rand & \rand & \rand \\

Ref.~\cite{TikhonovPRB18} & spin & $\alpha{=}\beta \in (d,2d)$ & \rand & \rand & \rand \\
Ref.~\cite{SchiroPRR20} & fermion & $\alpha,\beta \in (0,2.5)$ & \rand & \rand & \rand \\
Ref.~\cite{DetomasiPRB19} & fermion & $\alpha\in (1,4), \beta{=}\infty$ & \rand & \nonrand & \rand \\
Ref.~\cite{GargPRB19} & fermion  & $\alpha{=}\beta \in (0.5,3)$ & \nonrand & \nonrand & \rand \\ 
Ref.~\cite{NandKishorePRA19} & spin  & $\alpha{=}\beta\in (0.5,2.5)$ & \nonrand & \nonrand & \rand \\ 
Ref.~\cite{BarlevPRB20} & spin  & $\alpha{=}\beta \in (1.75,\infty)$ & \nonrand & \nonrand & \rand \\ 
Ref.~\cite{NagPRB20} & fermion & $\alpha\in (1.2,4), \beta{=}\infty$ & \nonrand & \nonrand & \rand\\
Ref.~\cite{RoySciPost19} & spin & $\alpha, \beta \in (0,\infty)$ & \nonrand & \nonrand & \rand  \\
Ref.~\cite{SantosPRL20} & boson & $\alpha, \beta \in ( 0,\infty)$ & \nonrand & \nonrand & \rand  \\\hline 
\end{tabular} \label{tab}
\caption{
Literature survey on models connecting to our model Eq. \eqref{e2} including fermion, boson and spin types. 
Disorder may enter in a hopping term,  $t_{ij}{=}\tilde{t}_{ij}/|i-j|^\alpha$, with randomized amplitude $\tilde t_{ij}$ and in a similarly randomized two-body density interaction,  $U_{ij}{=}\tilde{U}_{ij}/|i-j|^\beta$, and as randomized on-site potentials, $\epsilon_i$. 
In \cite{NandKishorePRA19,SantosPRL20} power-law correlated disorder has been implanted into $t_{ij}, U_{ij}$ by randomizing the site locations. 
Abbreviations: {\rand}  for randomized and {\nonrand}  for non-randomized; $\beta{=}\infty$ indicates nearest-neighbor interactions; SYK denotes an interaction as in Eq. \eqref{e1}, i.e. deviating from the density-density form. }
\label{t1} 
\end{table}
\section{Summary and Conclusion}
Inspired by the SYK-model we have defined a fermionic Hamiltonian with a one body term and a density-density type interaction, both fully randomized. Our conceptual motivation was to investigate the phase diagram of the model, which is interesting in the broader context of many-body localization: it exhibits a rich structure allowing for two integrable and several  chaotic phases, the latter being either ergodic or non-ergodic with respect to wavefunction spreading in Fock-space. Our practical motivation was that as compared to the SYK-model, our model is closer to microscopic realism in the sense that its interaction is of the density-density type. It therefore might help for identifying condensed-matter systems that exhibit phenomena to be interpreted in the SYK context. 

The findings of our computational analysis summarize as follows: 
(i) Our model exhibits two integrable points. While both exhibit a Poissonian level spacing statistic, they exhibit markably different spectral form factors.  
(ii) We find an RMT-type central phase that we argue to be adiabatically conntected with the RMT-phase of the original model. In particular, we observe a conformal scaling $G(\omega) \propto\omega^{-1/2}$ of the Green's function - at least within the numerical limitations of our study. 
(iii) Similar to an SYK-derived model studied in Refs. \cite{micklitz2019nonergodic,monteiro2020minimal}, our model also exhibits intermediate phases with RMT-type level spacing statistics and extended, but non-ergodic wavefunctions. In this regime the scaling of the IPR with the model parameters is quadratic, $(t/U)^2$, consistent with what was found numerically and analytically in Ref. \cite{monteiro2020minimal}.  
Summarizing, the model we propose represents a significant step from the SYK-model towards microscopic realism. It exhibits a phase diagram much richer as compared to SYK, but it still keeps a central phase that appears to exhibit all key features of SYK-physics.  

We conclude by putting our work into perspective with other studies from the recent literature; table \ref{t1} offers a brief account of relevant works. A variety of disorderd models has been studied limiting the range of interactions down to power-law envelopes for the kinetic (exponent $\alpha$) and interaction (exponent $\beta$) terms. They constitute links in an evolutionary chain born at the SYK model proper \eqref{e1}
and its variations in Tab. \ref{t1}, top section, and then connecting via our model Eq. \eqref{e2} followed by those listed in Tab. \ref{t1}, bottom section, to physical realism. How the corresponding phase diagram will evolve as moving along this chain is an interesting question, which here we have to leave to future research.

\section*{Acknowledgements}
FE thanks Alexander Altland for an inspiring conversation that has partially initiated this work. JD thanks Torsten Weber for insightful discussions. JD and FE acknowledge fruitful discussions on computational aspects with Christoph Lehner. SB received support from SERB-DST, India, through Ramanujan Fellowship (No. SB/S2/RJN-128/2016), Early Career Research Award (No. ECR/2018/000876), Matrics (No. MTR/2019/000566), and MPG for funding through the Max Planck Partner Group at IITB. Support from German Research Foundation (DFG) through the Collaborative Research Center, Project ID 314695032 SFB 1277 (projects A03, B01) and through EV30/11-1, EV30/12-1 and EV30/14-1 are  acknowledged.
\appendix

\section{Normalizing energy scale}
\label{appendix:norm}
\begin{figure}[t!]
    \centering
    \includegraphics[width=\linewidth]{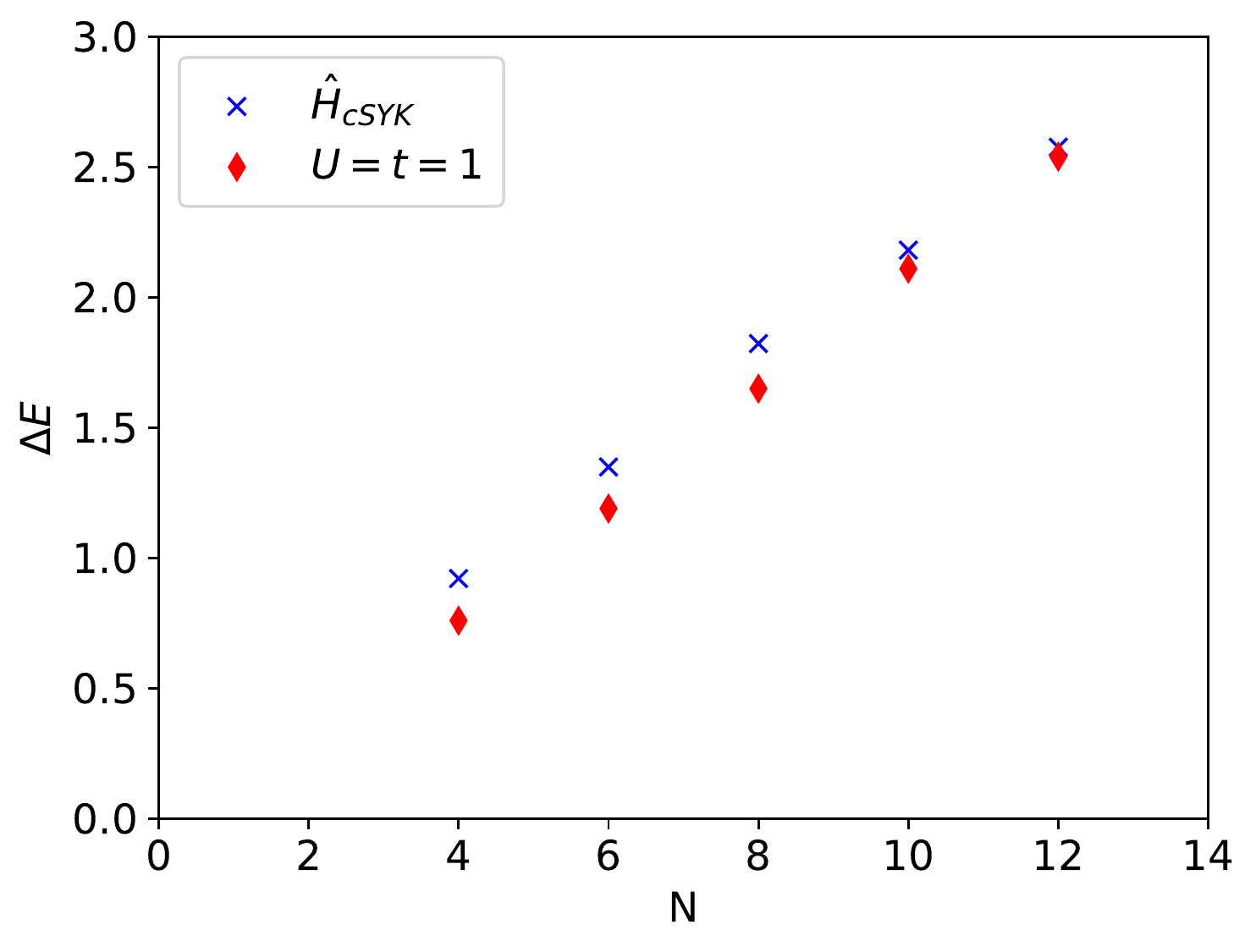}
    \caption{Extensive scaling of the bandwidth of the SYK model \eqref{e1} and Hamiltonian \eqref{e2}. We calculate $\Delta E$ for different system sizes $N_\mathcal{B}$. The bandwidth is averaged over $1000$ disorder realizations for $N_\mathcal{B}=4,6$ and $100$ realizations for $N_\mathcal{B}=8,10,12$. }
    \label{f7}
\end{figure}

In order to quantitatively compare results for the Hamiltonian models \eqref{e1} (parameter $J$) and \eqref{e2} (parameters $t,U$), we here determine the energy normalization.
We focus on the example $t=U$. 
The full bandwidth of the many-body problem $\Delta E=E_{max}-E_{min}$ scales extensively, see Fig. \ref{f7}, thus the normalization scale $\epsilon=\Delta E /N_\mathcal{B}$ is intensive. 
In order to compare the low frequency properties of the Green's function with the SYK model, we calculate the quantity $\epsilon$ for the SYK model and scale the  Hamiltonian \eqref{e2} so as to give the same value. For instance, in the case depicted in Fig. \ref{f7} we obtain $\epsilon \approx 0.22 J $. 
\\\\
\bibliographystyle{apsrev4-1}
\bibliography{paper.bib}

\begin{thebibliography}{58}
\bibitem [{foo({\natexlab{d}})}]{foot3}%
  \BibitemOpen
  \bibinfo {note} {Strictly speaking, RMT-statistics is expected only in the
  limit of large system sizes. However, as indicated by our data in Fig.
  \ref{f1} the asymptotics can be detected for this class of models already at
  relatively moderate system sizes.}\BibitemShut {Stop}%
\bibitem [{foo({\natexlab{e}})}]{foot4}%
  \BibitemOpen
  \bibinfo {note} {We here adopt the nomenclature $I_q$ for the IPR rather than
  the more conventional $P_q$ so as to avoid confusion with the level-spacing
  statistics $P(s)$.}\BibitemShut {Stop}%
\end{thebibliography}


\end{document}